\documentstyle[prl,aps,multicol,psfig]{revtex}

\begin{document}

\title{Phase diagram of mixed system of ferroelectric relaxors in the
random field theory framework}
\author{M.D.Glinchuk and E.A.Eliseev}
\address{Institute for Problems of Materials Science, Ukrainian NAS,
Krjijanovskogo str. 3, 03142, Kiev, Ukraine}
\author{V.A.Stephanovich}
\address{Institute of Mathematics University of
Opole, Oleska 48, 45-052 Opole, Poland \\ and
Institute of Semiconductor Physics NAS of Ukraine, Kiev, Ukraine}
\author{B.Hilczer}
\address{Institute of Molecular Physics,
PASc, Poznan, Poland}
\date{\today}
\maketitle

\begin{abstract}
We suggest a random field based model for calculation of physical properties 
of mixed ferroelectric relaxors. Our model naturally incorporates the 
different orientations of electric
dipoles (related to different solid solution components) as well as the 
contribution of nonlinear and correlation effects of random field.
We calculate the transition temperature $T_c$ as well as concentrational 
and temperature dependence of order parameters. 
The equations for these quantities have been derived. The theory has
been applied for description of the systems (PSN)$_{1-x}$(PST)$_x$ with different
degree of order as well as for (PMN)$_{1-x}$(PT)$_x$ systems. We show, that higher T$_c$ value
for more disordered (PSN)$_{1-x}$(PST)$_x$ system at 0 $\leq x<$
0.5 is related to larger nonlinearity coefficient of PSN in
comparison with that of PST. We determine the morphotropic region of 
temperatures and concentrations
for (PMN)$_{1-x}$(PT)$_x$. The observed phase diagram of
both aforementioned mixed ferroelectric relaxors is in pretty good coincidence
with the results of our numerical calculations.
\end{abstract}


\begin{multicols}{2}

\narrowtext

\section{Introduction}

The ferroelectric relaxors like PbMg$_{1/3}$Nb$_{2/3}$O$_3$ (PMN), PbSc$%
_{1/2}$Ta$_{1/2}$O$_3$ (PST), PbSc$_{1/2}$Nb$_{1/2}$O$_3$ (PSN), Pb$_{1-x}$La%
$_x$Zr$_{0.65}$Ti$_{0.35}$O$_3$ at $x$ = 0.08-0.09 (PLZT 8-9/65/35) have been
widely investigated in the last years [1]. In spite of this the mechanisms
of unusual properties of these materials (like Vogel-Fulcher (V-F) law in dynamic
dielectric permittivity, nonergodic behaviour, the distribution of different
properties maxima in broad temperature range ($\Delta T\sim $ 100 K for PMN
relatively to the temperature of permittivity maximum)) are still under
discussion. The presence of random field related to substitutional disorder, vacancies of
lead and oxygen, impurity atoms is firmly established in the relaxors.
Since this field has to influence the local properties of material, its
distribution was shown to result in V-F law [2], anomalies of nonlinear
dielectric permittivity [3], distribution of relaxation times and
non-Debye behaviour of dynamic dielectric permittivity [4]. The model [5]
for quantitative description of relaxor ferroelectrics is based
on the supposition that the random field destroys ferroelectric long range
order which could appear at $T\leq T_d$ where $T_d$ is Burns temperature ($%
T_d\sim $ 600 K for the most of relaxors). As a result, mixed ferroglass
phase with coexistence of long and short range order or dipole glass state
(both of them with non-ergodic behaviour) can appear. In the case of PLZT
the reference phase is PZT, its ferroelectric long range order is
destroyed completely (at $x_{La}=x\geq $ 0.07) or partly (at $x<$ 0.07) 
by random field of impurity La ions. The degree of disorder can be controlled by special
technological conditions which decrease the number of random field sources
in the relaxors PST, PSN and in other representatives of 1:1 family. In the
case of PMN (1:2 family relaxor) it appeared impossible to increase the
degree of order with the help of technological treatment. However, this can
be done by addition of PbTiO$_3$ (PT) (see e.g. [6] and ref. therein). In
particular, measurements had shown that in solid solution (PMN)$_{1-x}$(PT)$%
_x $ there is morphotropic phase transition between rhombohedral and
tetragonal phases at $x_c\approx $ 0.35. For $x>x_c$ normal ferroelectric
behaviour is present. Still missing theoretical description of this mixed
system phase diagram seems to be necessary to clarify the physical nature and
mechanisms of relaxor ferroelectrics behaviour. On the other hand the solution of these
problems is important due to giant electrostriction related to
high dielectric response which result in extremely high value of electromechanical
coefficient ($k\sim $ 0.92 for (PMN)$_{0.9}$(PT)$_{0.1}$ single
crystals). These anomalous properties made mixed ferroelectrics very
attractive for application in electromechanical transducers, actuators,
sonars etc. [7,8]. Another mixed system (PSN)$_{1-x}$(PST)$_x$ is also
prospective for aforementioned applications [9]. Both phase diagram of 
these relaxors and the behaviour of PSN and PST with
different degree of disorder exhibit the puzzles which have not been solved up to
now. The most interesting problem is the increase of transition temperature $T_c$
in PSN and its decrease in PST as disorder increases. The observed
dependence $T_c(x)$ at $0\leq x\leq 1 $ for solid solution of PSN and PST with
different degree of order had shown [10,11,12] that
aforementioned peculiarity exists in wide range of the components
concentrations, namely $T_c^{dis}(x)>T_c^{ord}(x)$ at $x\leq 0.5$.
The experimental and theoretical investigations of 
anomalies of dielectric response (the
existence of low and high temperature maxima obeyed to V-F and Arrhenius law
respectively) of PbSc$_{0.5}$(Nb$_{0.2}$Ta$%
_{0.3}$)O$_3$ single crystals 
have been performed recently [13]. In this work we propose the
method of calculation of mixed systems phase diagram. The method bases on the random field
concept. We calculate the concentrational dependencies of transition temperature and
order parameters. We show that the difference in PSN and PST behaviour is related to 
nonlinear and correlation effects contributions of random field. 
The calculated phase diagrams fit  
pretty good the measured ones for (PMN)$_{1-x}$(PT)$_x$ and (PSN)$_{1-x}$%
(PST)$_x$.

\section{The model}

Relaxors belong to the group of disordered ferroelectrics. In such materials
random site electric dipoles tend to order system via indirect dipole-dipole
interaction over soft mode of reference phase, while all other random
electric field sources try to disorder the materials. The competition
between these tendencies can lead to appearance of coherently oriented
dipoles. Their fraction $L$ can be calculated in the random field theory
framework on the base of equation

\begin{equation}
\vec{L}=\int\limits_{-\infty }^{+\infty }\left\langle
\vec{l}\right\rangle f(\vec{E},\vec{L})d%
\vec{E} \label{e1}
\end{equation}

where $f(E,L)$ is random field $E$ distribution function.

As the matter of fact $L$ is a dimensionless order parameter, which
expressed through single dipole moment $\left\langle \vec{l}%
\right\rangle =\frac{\left\langle \vec{d^{*}}\right\rangle }{%
d^{*}}$ ($d^{*}$ is effective dipole moment related to Lorenz field
parameter, see [14] for the details) averaged over possible orientations and
random field distribution function. In the case of two-orientation dipoles ($%
l_z=1,l_x=l_y=0$) $\langle l_z\rangle =\tanh(d^{*}E_z/kT$), Eq.(1) can be
rewritten as

\begin{equation}
L_z=\int \tanh \left( \frac{d^{*}E_z}{kT}\right) f(E_z,L_z)dE_z \label{e2}
\end{equation}

For the completely ordered system which is usually described by a mean field
approximation $f(E,L)=\delta (E-E_0L)$, where $E_0$ is mean field value ($%
d^{*}E_0=kT_{cmf}$), so that Eq.(2) gives well-known expression for
order-disorder ferroelectrics: $L_{mf}=\tanh (d^{*}E_0L_{mf}/kT)$ and $%
T\to T_{cmf}$ at $L_{mf}\to 0$.

For the considered case of mixed system of ferroelectric relaxors A$_{1-x}$B$%
_x$ two types of electric dipoles $d_{1z}^{*}$ and $d_{2z}^{*}$ and order
parameters $L_{1z}$ and $L_{2z}$ must be considered. In such a case the
random field is induced by both type of dipoles and its distribution
function $F$ depends on both order parameters, i.e. $F=f(E,L_{1z},L_{2z})$.
Therefore $L_{1z}$ and $L_{2z}$ can be written in the form of Eq.(2) with
this distribution function and with $d_{1z}^{*}$ and $d_{2z}^{*}$ in the
Eqs. for $L_{1z}$ and $L_{2z}$ respectively, $z$ and $z$ being the
directions of 1 and 2 types electric dipoles orientations.

The most of the disordered system characteristics are included into random
electric field distribution function $f(E,L)$. This distribution function
has been calculated in the statistical theory framework allowing for
contribution of electric dipoles, point charges and dilatational centers as
the random field sources [14]. Linear, nonlinear and spatial correlation
effects contribution of the random field was taken into account in [15].
Allowing for these effects for the materials with cubic symmetry (see [15])
in supposition that the considered electric dipoles are the main sources of
random field and introducing $\cos(l_1l_2)\equiv \cos(zz^{\prime })$ one can
obtain the following equations for two order parameters:
\end{multicols}
\widetext
\noindent\rule{20.5pc}{0.1mm}\rule{0.1mm}{1.5mm}\hfill
\begin{eqnarray}
L_1 \equiv L_{1z}=\int_{-\infty }^{\infty } \tanh \frac{%
d_{1z}^{*}(\varepsilon _z+\alpha _3^{(1)}\varepsilon _z^3)}{kT}%
f_1(\varepsilon _z,L_1,L_2)d\varepsilon _z,  \nonumber \\
L_2 \equiv L_{2z^{\prime }}=\int_{-\infty }^{\infty } \tanh \frac{%
d_{2z^{\prime }}^{*}(\varepsilon _{z^{\prime }}+\alpha _3^{(2)}\varepsilon
_{z^{\prime }}^3)}{kT}f_2(\varepsilon _{z^{\prime }},L_1,L_2)d\varepsilon
_{z^{\prime }}. \label{e3}
\end{eqnarray}

Here $\alpha _3^{(1,2)}$ is coefficient of nonlinearity of the reference
phase, its dimensionality being $\varepsilon ^{-2}$ (see [15]).

\begin{eqnarray}
f_1(\varepsilon _z,L_1,L_2) &=&\frac 12\int\limits_{-\infty }^{+\infty }\exp
(-\rho _z^2\eta _1) \cos \left[ \rho _z\left( \varepsilon _z-(1-x)\frac{T_{cmf2}}{d_2^{*}}%
L_2\cos (\widehat{l_1l_2})-x\frac{T_{cmf1}}{d_1^{*}}L_1\right) \right] d\rho
_z, \label{e4} \\
f_2(\varepsilon _{z^{\prime }},L_1,L_2) &=&\frac 12\int\limits_{-\infty
}^{+\infty }\exp (-\rho _{z^{\prime }}^2\eta _2) 
\cos \left[ \rho _{z^{\prime }}\left( \varepsilon _{z^{\prime }}-(1-x)%
\frac{T_{cmf2}}{d_2^{*}}L_2-x\frac{T_{cmf1}}{d_1^{*}}L_1\cos (\widehat{l_1l_2%
})\right) \right] d\rho _{z^{\prime }},  \nonumber \\
\eta _1 &=&\cos ^2(l_1l_2)\xi _1+\xi _2;\ \eta _2=\xi _1+\cos
^2(l_1l_2)\xi _2; \text{ }
\xi _1 =\frac{16\pi }{15}\frac{d_1^{*2}}{\varepsilon _1^2r_{c1}^3}n_1;%
\text{ }\xi _2=\frac{16\pi }{15}\frac{d_2^{*2}}{\varepsilon _2^2r_{c2}^3}n_2.
\nonumber
\end{eqnarray}
\hfill\rule[-1.5mm]{0.1mm}{1.5mm}\rule{20.5pc}{0.1mm}
\begin{multicols}{2}
\narrowtext
\noindent
In Eq.(\ref{e4}) $\xi _1$ and $\xi _2$ describe the width of random field
distribution functions with Gaussian form related, respectively, to the first
and second types of dipoles, their concentrations being $n_1=\frac{\beta _1x%
}{a_1^3}$ and $n_2=\frac{\beta _2(1-x)}{a_2^3}$ ($\beta _i$ is fraction of
unit cells in which dipoles exist); $\varepsilon _{1,2}$, $a_{1,2}$ and $%
r_{c1,2}$ are respectively dielectric permittivity, lattice constants and
correlation radii of A and B relaxors reference phase. Note, that the
difference between distribution functions $f_1$ and $f_2$ is related to the
different orientations of the dipoles which manifests itself in the presence
of $\cos (l_1l_2)$ multiplier; $f_1=f_2$ if $\cos (l_1l_2) =1$.

The polarization of mixed system $P$ can be expressed via $L_1$ and $L_2$ as
follows
\begin{equation}
\vec{P}=x\frac{\beta _1\vec{d_1^{*}}L_1}{a_1^3}+(1-x)%
\frac{\beta _2\vec{d_2^{*}}L_2}{a_2^3}.\label{e5}
\end{equation}

\section{Phase diagram of mixed system}

The phase diagram has to describe the dependence of the temperature $T_c$ at
which order parameters arise on concentration. The $T_c(x)$ dependence can
be obtained from Eqs.(\ref{e3}), (\ref{e4}) in the limit $L_1 \to 0$, $L_2 \to 0$. One can see
from (\ref{e4}), that in such a limit the distribution functions $f_1$ and $f_2$
can be represented in the form:

\begin{eqnarray}
&&f_{1,2}(\varepsilon _i,L_1,L_2) =f_{01,2}(\varepsilon _i)-k_{1,2}\left(
\frac{df_{1,2}}{d\varepsilon _i}\right) _{L_1=L_2=0},\nonumber \\
&&i=z,z^{\prime }, \label{7} \\
&&k_1 =(1-x)\frac{T_{cmf2}L_2}{d_2^{*}}\cos (\widehat{l_1l_2})+x\frac{%
T_{cmf1}L_1}{d_1^{*}};  \nonumber \\
&&k_2 =(1-x)\frac{T_{cmf2}L_2}{d_2^{*}}+x\frac{T_{cmf1}L_1}{d_1^{*}}\cos (%
\widehat{l_1l_2}),  \label{8} \\
&&f_{01,2}(\varepsilon _i) =\frac 1{2\sqrt{\pi \eta _{1,2}}}\exp \left( -%
\frac{\varepsilon _i^2}{4\eta _{1,2}}\right) .  \label{9}
\end{eqnarray}

Substituting Eqs. (\ref{7})-(\ref{9}) into (\ref{e3}) and performing the integration, one
obtains the following system of equations for $T_c(x)$ calculation:

\begin{eqnarray}
&&L_1 =\left[ (1-x)T_{cmf2}L_2\Delta \cos (l_1l_2)+xT_{cmf1}L_1\right] \frac{%
I_1}{kT_c};  \nonumber \\
&&L_2 =\left[ (1-x)T_{cmf2}L_2+x\frac{T_{cmf1}}\Delta L_1\cos
(l_1l_2)\right] \frac{I_2}{kT_c};   \label{10} \\
&&I_{1,2} =\int\limits_{-\infty }^{\infty }f_{01,2}(\varepsilon )\frac{%
(1+3\varepsilon ^2\alpha _3^{(1,2)})d\varepsilon }{\cosh^2(d_{1,2}^{*}/kT_c)(%
\varepsilon +\alpha _3^{(1,2)}\varepsilon ^3)};\nonumber \\
&&\Delta =\frac{d_1^{*}}{d_2^{*}}.    \label{11}
\end{eqnarray}

Integrals $I_1$ and $I_2$ depend on components concentrations since the
distribution function width (see Eq. (\ref{9})) depends on $x$ and $T_c$.
\begin{figure}[tbh]
\vspace*{-4mm}
\centerline{\centerline{\psfig{figure=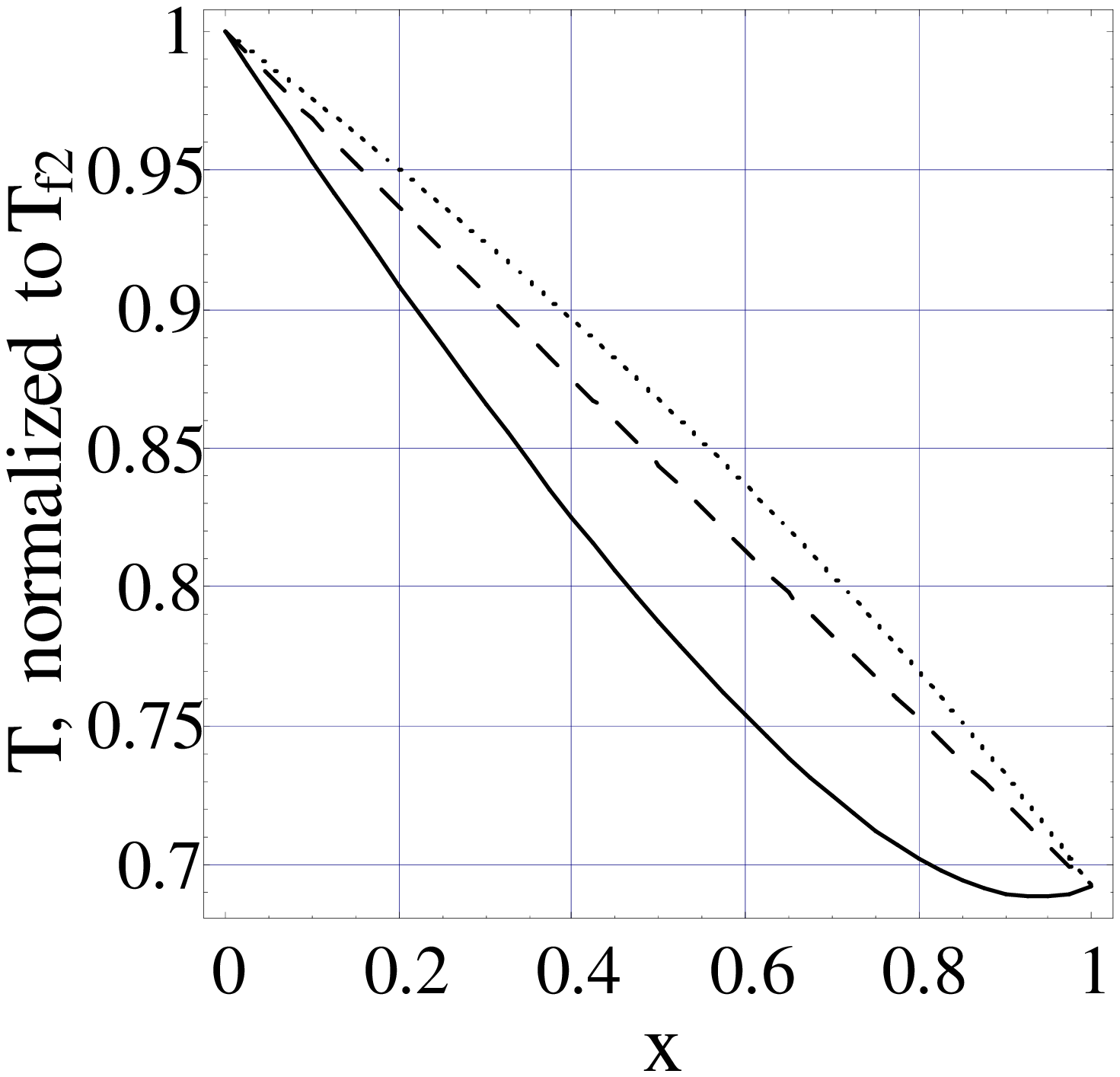,width=0.8\columnwidth}}}
\vspace*{-8mm}
\center {\text{a}}
\vspace*{-10mm}
\centerline{\centerline{\psfig{figure=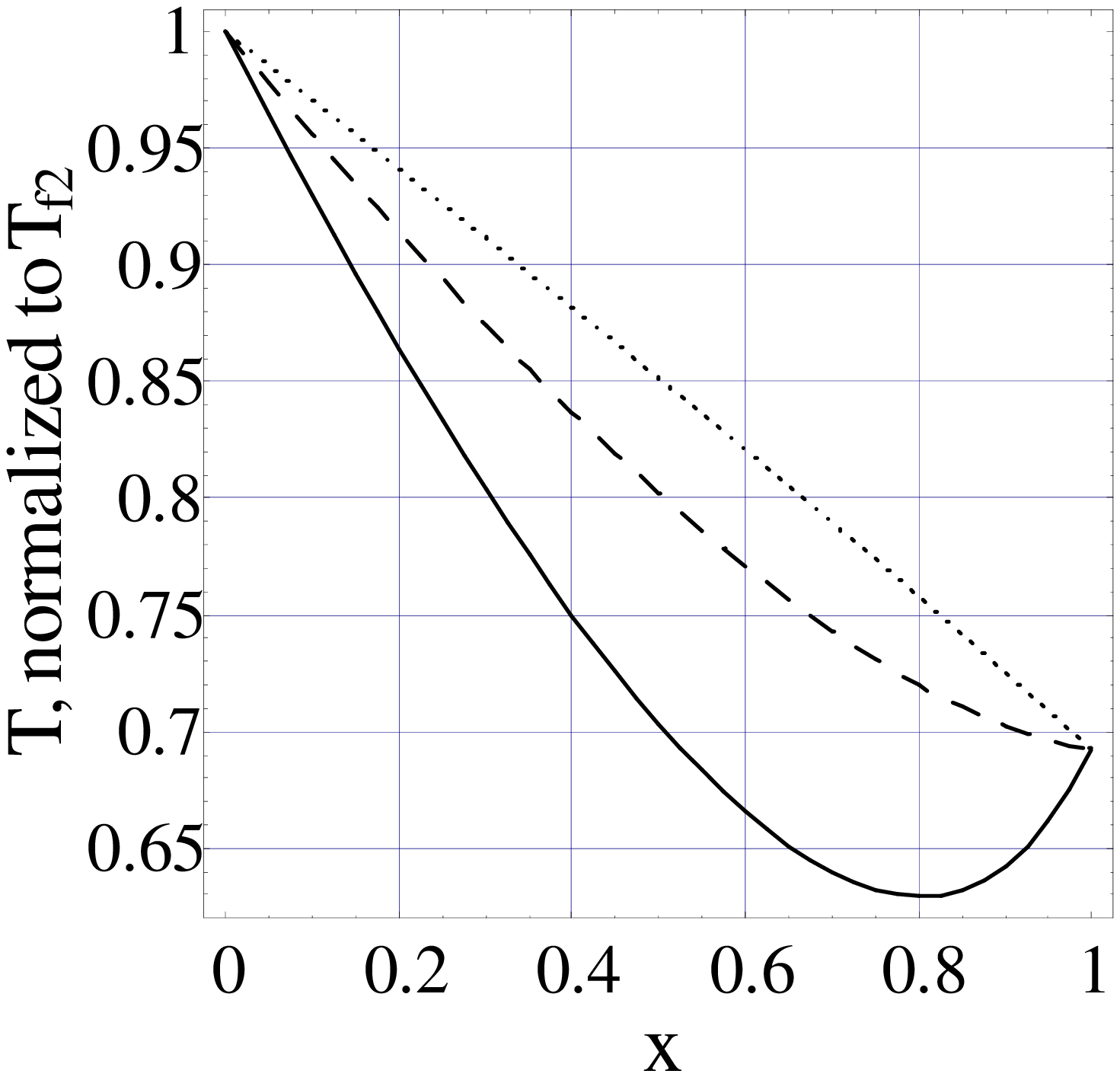,width=0.8\columnwidth}}}
\vspace*{-8mm}
\center { \text{b}}
\vspace*{1mm}
\caption{Concentrational dependence of dimensionless transition temperature $%
\tau_c = \frac{T_c(x)}{T_{cmf2}}$ for the case $d_1 \parallel d_2$, $\frac{%
d_2^*}{d_1^*}=0.6$ (solid line); 0.7 (dashed line); 0.8 (dotted line); $%
\alpha_0^{(1)}=\alpha_0^{(2)}=1$ (a); $\alpha_0^{(1)}=\alpha_0^{(2)}=3$ (b).
}
\end{figure}
The condition of the system (\ref{10}) solvability leads to following
equation for $T_c$:
\begin{eqnarray}
&&T_c^2-C_2(T_c)T_c+C_1(T_c)=0;   \label{12} \\
&&C_1(T_c) =(1-x)xI_1I_2T_{cmf1}T_{cmf2}(1-\cos ^2(\widehat{l_1l_2}));
\nonumber \\
&&C_2(T_c) =xI_1T_{cmf1}+(1-x)I_2T_{cmf2}.   \label{13}
\end{eqnarray}
The dependencies $C_{1,2}(T_c)$ are related to those of $I_{1,2}$ integrals
(see Eqs.(\ref{10})), so that (\ref{13}) is complicated nonlinear equation 
for $T_c$. It will be solved numerically.

Factor $(1-\cos ^2(l_1l_2))$ reflects the dependence on orientations of
the $d_1^{*}$ and $d_2^{*}$ dipoles. It equals $\frac 23$ in the
case when the dipoles 1 and 2 are oriented, respectively, along [001] and
[111] type of direction. This factor is zero as $\cos (l_1l_2)%
=\pm 1 $ when the dipoles are parallel or antiparallel to each other.
In the latter case the system (\ref{13}) reduces to single equation 
$T_c=C_2$.
One can see, that in general case there is at least two $T_c$ values and the
largest one has to describe the mixed system behaviour. It is seen from Eqs.
(\ref{12}), (\ref{13}), (\ref{11}), (\ref{9}), that $T_c$ depends on 
several parameters - coefficients
of nonlinearity, distribution function half- width and electric dipole
moments ratio. To illustrate the influence of these parameters on $T_c(x)$,
we depicted in Fig.1 the dimensionless transition temperature $\tau
_c=T_c(x)/T_{cmf2}$ for the case when $d_1\parallel d_2$ (i.e. $T_c=C_2$)
for several values of dimensionless parameters. It is seen that
increase of $\alpha _0^{(1,2)}$ and $\Delta =d_1^{*}/d_2^{*}$ increases the rate of $T_c(x)$
decrease. The influence of dipole moments ratio is more pronounced 
with decrease of the second component concentration.
\begin{figure}[tbh]
\vspace*{-4mm}
\centerline{\centerline{\psfig{figure=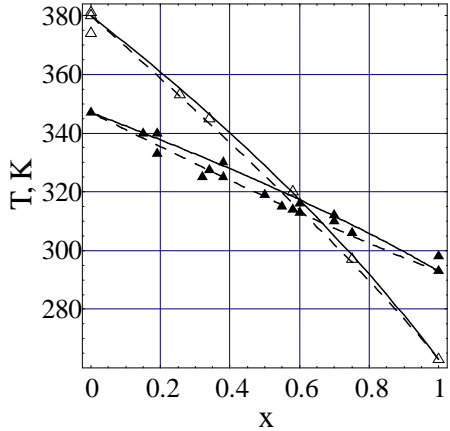,width=0.8\columnwidth}}}
\vspace*{-3mm}
\caption{Phase diagram of mixed relaxors (PSN)$_{1-x}$(PST)$_x$ for
disordered (open symbols) and ordered (solid symbols) ceramic samples, taken
from [10,11,12]. Solid and dashed lines - theory respectively for
aforementioned and 10\% larger $\Delta$ value with $\alpha_0^{(1)}=0$, $%
\alpha_0^{(2)}=0,3$ and $\alpha_0^{(1)} =\alpha_0^{(2)}=0$ for disordered
and ordered materials.}
\end{figure}

\section{Mixed system of 1:1 family relaxors: (PSN)$_{1-x}$(PST)$_x$}

The behaviour of this solid solution components (PSN and PST) is strongly
different: the increase of the degree of disorder leads to increase of $%
T_c$ for PSN and to decrease of $T_c$ for PST. At the first glance the
latter seems to be more reasonable while the PSN behaviour is puzzling
since it is common wisdom that random field decreases $T_c$. On the other
hand the increase of random field value can result into appearance of nonlinear
and correlation effects. These effects have to be dependent on the value of
nonlinearity coefficient of the material. Keeping in mind that PSN and PST
contain scandium, niobium and tantalum oxygen complexes, one can
conclude that coefficient of nonlinearity for PSN should be larger than that for
PST. This is because this coefficient for niobium oxygen complexes was shown
to be several times larger than that for tantalum [16]. The value
of nonlinearity coefficients for PSN and PST was extracted from observed $%
T_c $ values with the help of formulas (\ref{12}), (\ref{13}) at $x=0$ (PSN) and $x=1$
(PST). One can see, that in both cases $C_1(T_c)=0$ and

\begin{eqnarray}
T_c(PSN) &=&I_2T_{cmf2},   \label{14} \\
T_c(PST) &=&I_1T_{cmf1}.  \nonumber
\end{eqnarray}
\begin{figure}[tbh]
\vspace*{-4mm}
\centerline{\centerline{\psfig{figure=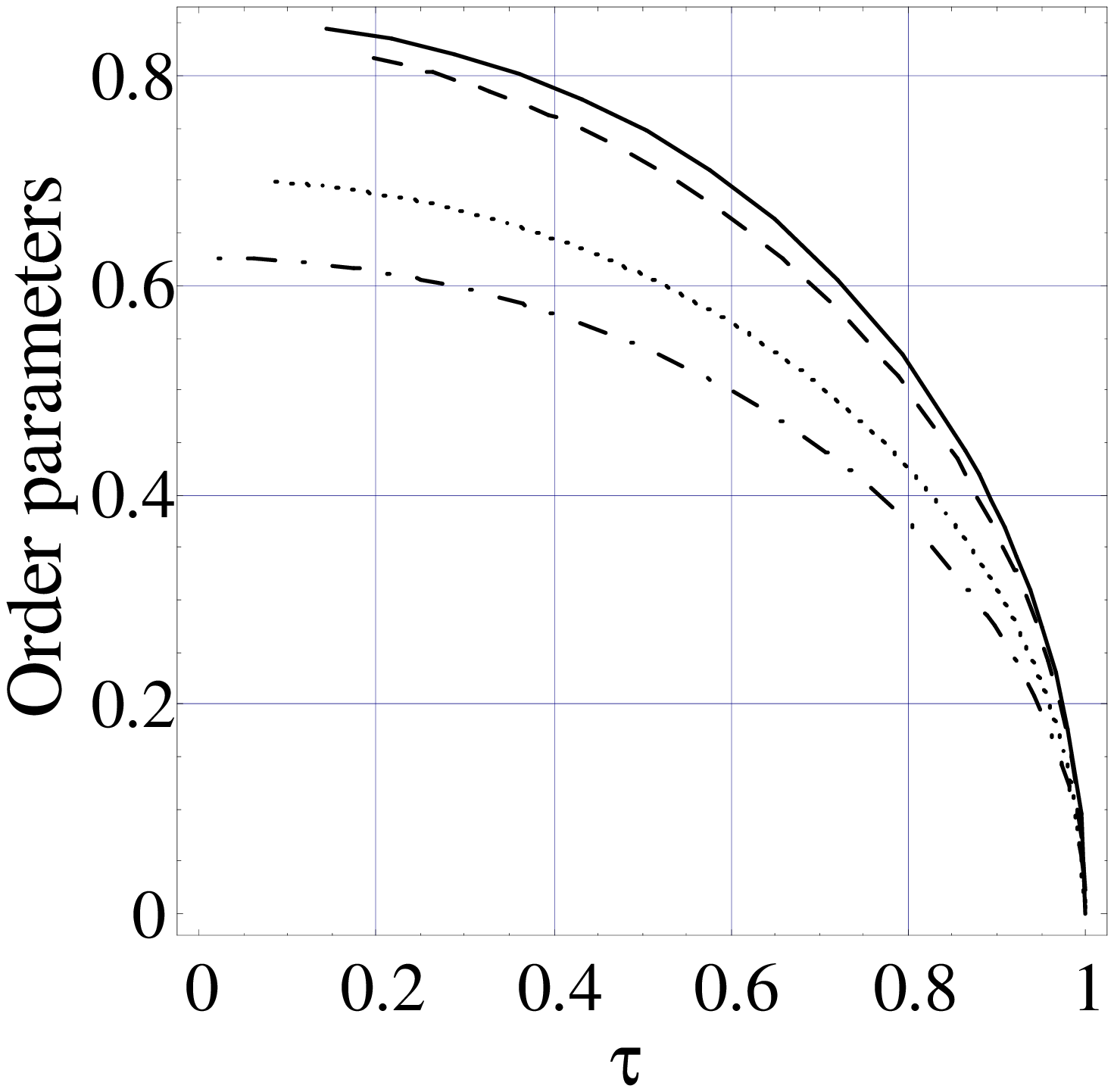,width=0.8\columnwidth}}}
\vspace*{-10mm}
\center {\text{a}}
\vspace*{-10mm}
\centerline{\centerline{\psfig{figure=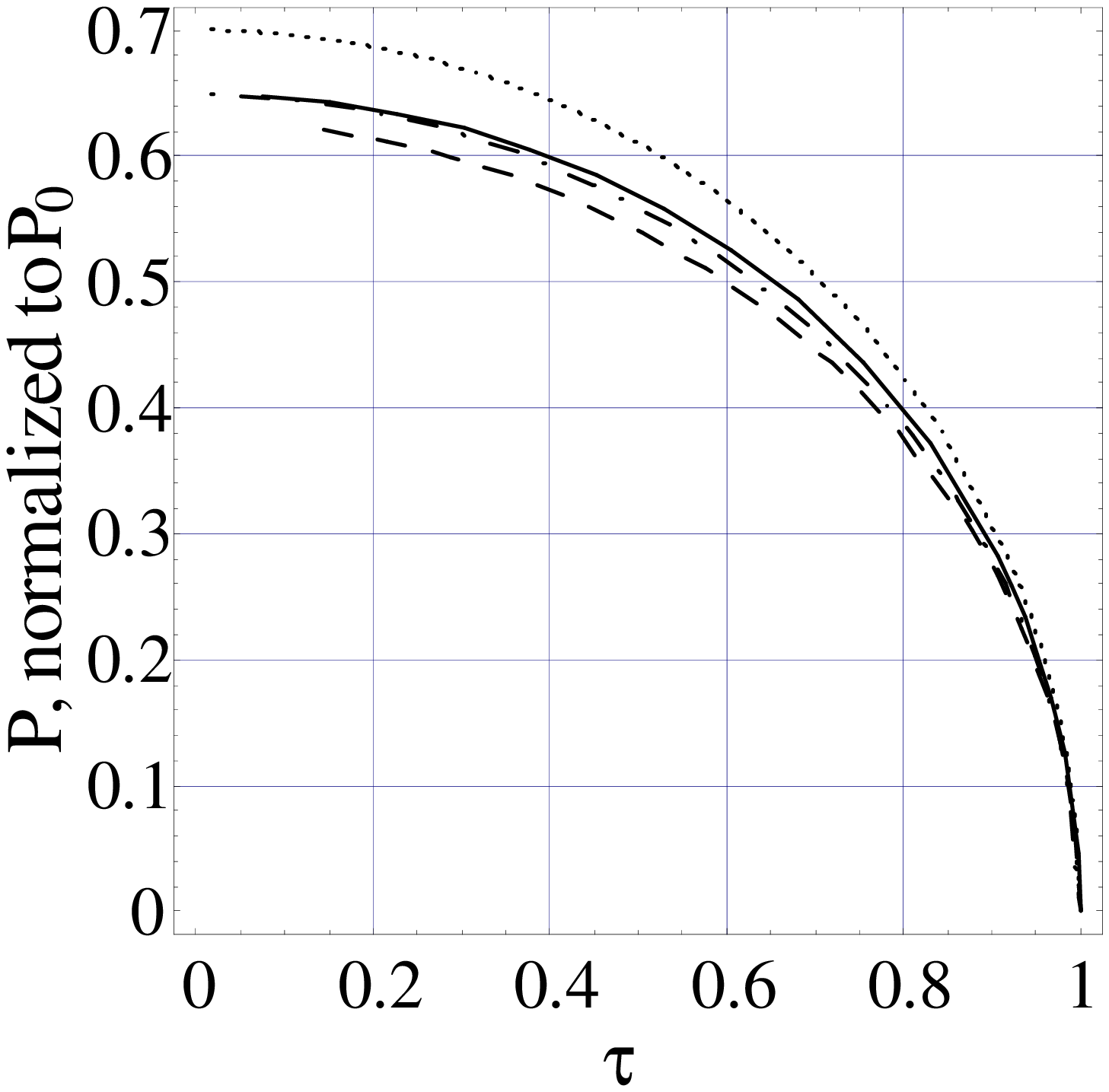,width=0.8\columnwidth}}}
\center { \text{b}}
\vspace*{3mm}
\caption{Fraction of coherently ordered dipoles in (PSN)$_{1-x}$(PST)$_x$.%
\newline
a) solid line - $L_{2ord}(x=0)$; dashed line - $L_{2disord}(x=0)$; dotted
line - $L_{1ord}(x=1)$; dotted-dashed line - $L_{1disord}(x=1)$ \newline
b) solid line - $P_{ord}(x=0,3)$; dashed line - $P_{disord}(x=0,3)$; dotted
line - $P_{ord}(x=0,8)$; dotted-dashed line - $P_{disord}(x=0,8)$. }
\end{figure}
The fitting of observed $T_c$ values (see e.g. [10] and ref. therein) for
more disordered PSN and PST made it possible to obtain dimensionless
coefficient of nonlinearity ($\alpha _0$ $=\alpha _3(kT_{cmf2}/d_1^{*})$)
for PSN $\alpha _0^{(2)}=0,3$ and $\alpha _0^{(1)}=0,098$ for PST along with
distribution function dimensionless half-width ($q_{1,2}=\frac{\sqrt{\xi _0}%
d_{1,2}^{*}}{kT_{cmf1,2}},$ $\xi _0=\frac{16\pi }{15}\frac{d_1^{*2}}{%
\varepsilon ^2r_c^3}\frac \beta {a^3}$ ) $q_2$(PSN) = 0.425, $q_1$(PST) =
0.499. Note, that while introducing dimensionless values we took into account,
that parameters of reference phase for PSN and PST are close to each other.
Allowing for the fact that $T_{cmf}$(PSN) $\approx T_{cmf}$(PST) [17] one comes to
conclusion that $\frac{q_1}{q_2}\approx \overline
{d_1^{*}}{d_2^{*}}^2=\Delta ^2$ which gives $\Delta ^{-1}=0.89$ for more
disordered relaxors. The same fitting of $T_c$ values observed in more
ordered PSN and PST [10,11] leads to $\alpha _0^{(1)}=\alpha _0^{(2)}\approx
0$, $q_1=0.477$, $q_2=0.416$ and $\Delta ^{-1}=0.88$. The obtained
parameters made it possible to calculate $T_c(x)$ both for more disordered
and ordered (PSN)$_{1-x}$(PST)$_x$ mixed system without additional fitting
parameters. To simplify the calculations we supposed the same orientations
of $d_1$ and $d_2$ dipoles. The comparison of calculated and measured
[10-12] $T_c(x)$ values is shown in Fig.2. We used two different values of $\Delta $
to show, that the concentration $x_c$ at which $T_c$(disordered) = $%
T_c$(ordered) is sensitive to the parameter values. 
One can see that the
theory describes the observed values pretty good, allowing for the
distribution of measured $T_c$ values related to different experimental
conditions (see [10] and ref. therein). 

The calculations of $L_1$ and $L_2$
on the base of Eq.(\ref{e3}) had shown, that $L_2>L_1$ both for more disordered and
ordered relaxors, the values of $L$ for ordered ones being larger than those
for more disordered samples. The nonzero values of $L_1$ and $L_2$ in whole
range of the components concentrations give evidence that each component
is in mixed ferro-glass phase (see also [13]).
\begin{figure}[tbh]
\vspace*{-4mm}
\centerline{\centerline{\psfig{figure=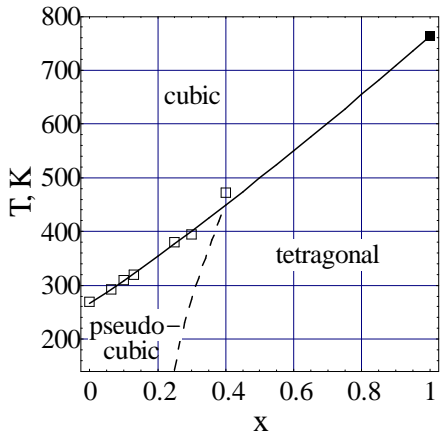,width=0.8\columnwidth}}}
\caption{Phase diagram of (PMN)$_{1-x}$(PT)$_x$. Experimental data [6] -
symbols; theory - solid line is $T_c (x)$ and dashed line is morphotropic
region.}
\end{figure}

The results of numerical calculations for several concentrations of 
solid solution components
is depicted in Fig.3. In Fig.3a we
represented the fraction of coherently ordered dipoles in PSN ($x=0$) and
PST ($x=1$) which is proportional to these materials polarization. For two
relaxors solid solution at $x=0.3$ and $x=0.8$ we performed the calculations
of dimensionless polarization (in the units of $P_0=\frac{d_1^{*}}{%
a_1^3}$) on the base of Eq.(\ref{7}) (keeping in mind that $d_1\parallel d_2$ and $%
\beta _1=\beta _2$) and showed the results in Fig.3b. It is seen that
increase of $x$ leads to increase of polarization, $P(x)$ of more ordered samples is
larger than $P(x)$ of disordered samples.

\section{Mixed system (PMN)$_{1-x}$(PT)$_x$}

This system is a solid solution of relaxor ferroelectric PMN (component 2)
and normal ferroelectric PbTiO$_3$ (component 1). In fact, the
phase diagram of this system can be calculated similar to that
in the previous section, with some details different.

First, due to the existence of normal ferroelectric phase transition in
PbTiO$_3$ at $T_c=763$ K which can be calculated in a mean field
approximation, $T_{cmf1}=763$ K. Second, the distribution function of
random field in ordinary ferroelectrics is known to be $\delta $-function
because of negligibly small random field in the system, so that $\xi _1=0$
in Eq.(\ref{e4}). Note, that in such a case PT component contributes to random
field distribution via PT mean field written as $xT_{cmf1}L_1/d_1^{*}$ (see
Eqs.(\ref{e4})). The number of electric dipoles $n_1=x/a_3$ where $a$ is PT lattice
constant. 
\begin{figure}[tbh]
\vspace*{-4mm}
\centerline{\centerline{\psfig{figure=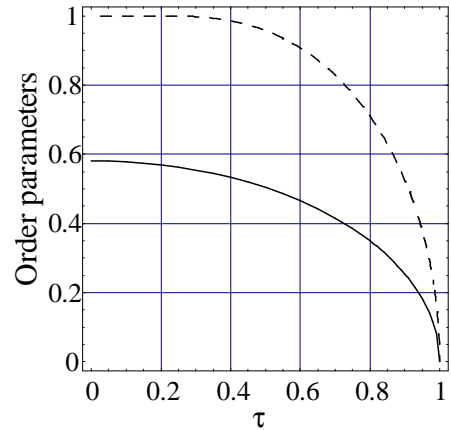,width=0.8\columnwidth}}}
\caption{Temperature dependence of the fraction of coherently ordered
dipoles in PMN (solid line) and PT (dashed line).}
\end{figure}
It is important to emphasize that PMN relaxor state was considered
as mixed ferroglass phase with coexistence of long and short range order.
The peculiarities of nonlinear dielectric permittivity of PMN (see [18])
speak in favour of this statement. 
Keeping in mind that the dipoles in PMN
and PT are oriented, respectively, along [111] and [100] type of directions, we
calculated $T_c(x)$ on the base of Eqs.(\ref{12}), (\ref{13}) with $\cos (l_1l_2)=\frac
1{\sqrt{3}}$. It appeared possible to neglect the contributions of nonlinear
and correlation effects because they do not improve the fitting of
calculated and measured $T_c(x)$ dependence. 
The results of numerical
calculations are reported in Fig.4. Fitting parameters $q_2=0.51$ and $%
\Delta =\frac 13$ were obtained from observed $T_c$(PMN), $T_c$%
(PT) and $\frac{\beta _2}{\beta _1}\frac 1\Delta =\frac 13$, i.e. $\frac{%
\beta _2}{\beta _1}=\frac 19$, so that only 11\% of unit cells in PMN have
electric dipoles because $\beta _1 =1$. One can see that our calculation
describes pretty good observed $T_c(x)$ dependence.

To find the change of mixed system symmetry with variation of concentration of
components, we performed the calculations of $L_1$ and $L_2$ 
temperature and concentrational dependence. We start with calculations for
PMN ($x=0$) and PT ($x=1$) using the aforementioned parameters (see Fig.5).
The comparison of solid and dashed lines in Fig.5 shows clearly the
difference of the behaviour of order parameters in mixed ferroglass phase
(PMN) and normal ferroelectric phase (PT). The calculation of $L_2(T)$ at $%
T\to 0$ had shown that $L_2(T=0)\approx 0.58$, so that the
contribution of long range order is large enough in low temperature region
in PMN. 
\begin{figure}[tbh]
\vspace*{-4mm}
\centerline{\centerline{\psfig{figure=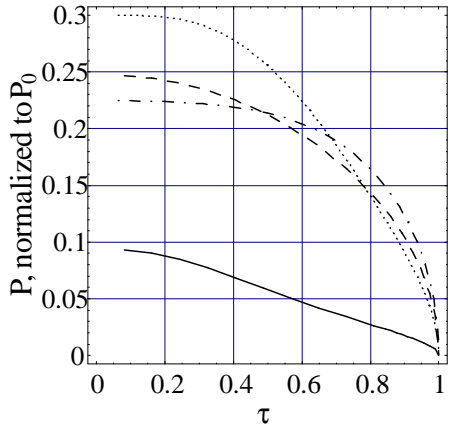,width=0.8\columnwidth}}}
\caption{Temperature dependence of the components with rhombic ($P_2$) and
tetragonal ($P_1$) symmetry of dimensionless polarization in (PMN)$_{1-x}$%
(PT)$_x$. Solid line - $P_2(x=0,1)$; dotted line - $P_1(x=0.3)$; dashed line
- $P_1(x=0.1)$; dotted-dashed line - $P_2(x=0.3)$.}
\end{figure}
The polarization of mixed system was calculated with the help
of Eq.(6), which incorporates the contribution of the first and second type
of dipoles, i.e. $\vec{P}(x)=\vec{P_1}(x)+%
\vec{P_2}(x)$. The concentrations at which $\left|
\vec{P_1}(x)\right| =\left| \vec{P_2}(x)\right| $
correspond to coexistence of two types of symmetry in the mixed system, i.e.
to appearance of morphotropic region in phase diagram. Temperature
dependence of dimensionless $\left| \vec{P_1}\right| $ and $%
\left| \vec{P_2}\right| $ for several $x$ values is depicted in
Fig.6.

The most interesting feature in the Fig.6 is the existence of the curves
crossover at some $T=T_{cr}(x)$ for $x=0.3$. This means that $\left|
P_1\right| =\left| P_2\right| $ at $T=T_{cr}(x)$ (morphotropic region), $%
\left| P_1\right| >\left| P_2\right| $ at $T>T_{cr}(x)$ or $\left|
P_1\right| <\left| P_2\right| $ at $T<T_{cr}(x)$, i.e. mixed system has the
symmetry of component 1 or component 2 respectively. For (PMN)$_{1-x}$(PT)$%
_x $ the value $T_{cr}(x=0.3)=275$ K and $T_{cr}(x=0.4)=460$ K, therefore $%
T_{cr}$ increase with concentration increase (see dashed line in Fig.4). On
the base of this consideration we discern the symmetry of mixed (PMN)$_{1-x}$(PT)$%
_x$ system in Fig.4 (pseudo-cubic like PMN and tetragonal like PT) which is
in agreement with observed one [6].

One can see from Fig.6 the absence of the curves crossover, i.e. morphotropic
region at $x=0.1$. Since the compositions with small concentration of PT ($%
x\approx 0.1$) are important for applications (see [7,8]) we depicted in
Fig.7 the fraction of coherently ordered dipoles and the components of
polarization with tetragonal (solid line) and pseudo-cubic (rhombohedral)
symmetry (dashed line) at $T=0$. 
\begin{figure}[tbh]
\vspace*{-4mm}
\centerline{\centerline{\psfig{figure=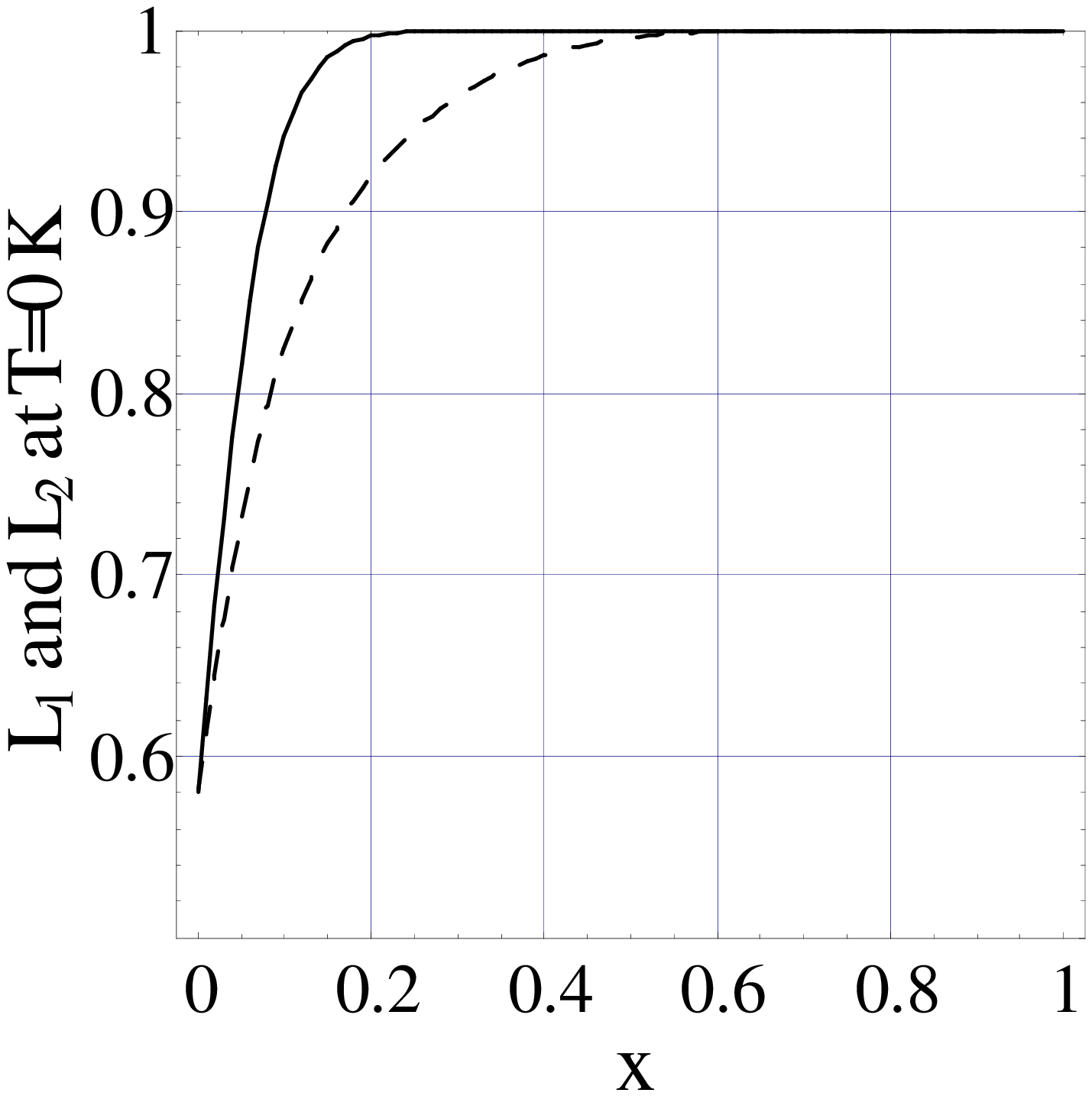,width=0.8\columnwidth}}}
\vspace*{-8mm}
\center {\text{a}}
\vspace*{-10mm}
\centerline{\centerline{\psfig{figure=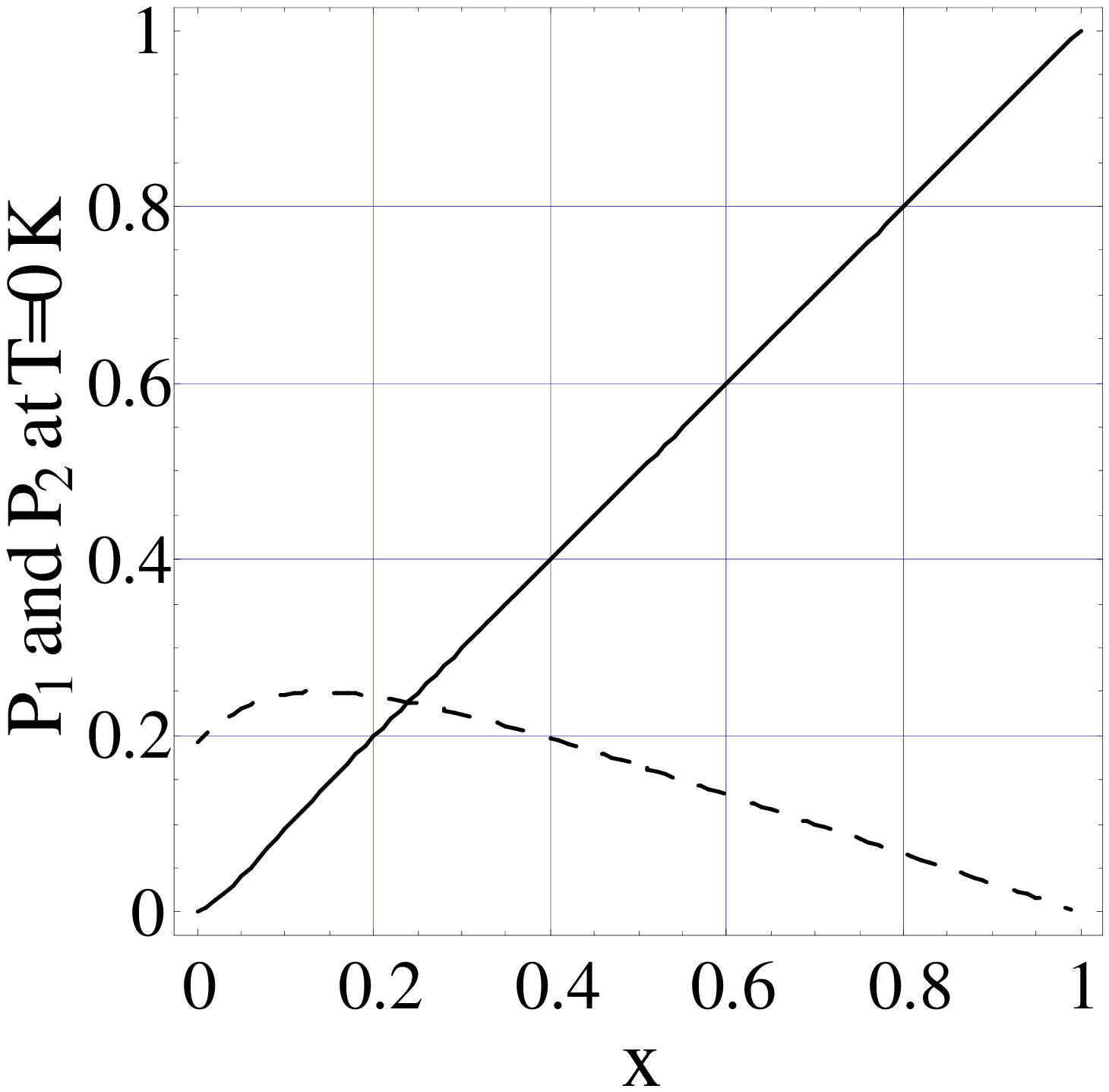,width=0.8\columnwidth}}}
\center { \text{b}}
\vspace*{3mm}
\caption{Concentrational dependence of order parameters at $T=0$ with
pseudo-cubic (rhombohedral) (dashed lines) and tetragonal (solid lines)
symmetry expressed via fraction of coherently ordered dipoles (a) and via
contributions to polarization (b).}
\end{figure}
It is seen from Fig.7b that tetragonal
component linearly increases with PT concentration increase whereas
pseudo-cubic component has a maximum at $x\approx 0.135$. 
This maximum
origin can be related to the competition between increase of $L_2$ (see
dashed line in Fig.7a) and $(1-x)$ decrease with $x$ increase (see Eq.(6)).
Because of pseudo-cubic symmetry, both the polarization maximum and maxima of
dielectric response, piezoelectric and electromechanical coefficient can be
expected at $x=0.135$. This concentration is a little larger than $x\approx
0.1$ where the high values of aforementioned properties were observed [7,8].

\section{Conclusion}

We propose a model for calculation of ferroelectric order parameters
and phase diagram of mixed relaxors. The physical background of the model is
influence of random electric field of mixed system on its properties.
Randomly distributed electric dipoles were supposed to be the main sources
of the field. The contribution of nonlinear and spatial
correlation effects of random field was taken into account.
We carried out the specific calculations for mixed PSN-PST and
PMN-PT systems. The solution of the puzzle of larger transition
temperature of more disordered (PSN)$_{1-x}$(PST)$_x$ system in
the region $0\leq x<0,5$ was shown to be related to nonlinear and
correlation effects contribution which has to be larger in PSN.
The obtained concentrational
dependence of transition temperature and polarization of solid solution (PMN)%
$_{1-x}$(PT)$_x$ revealed the existence of morphotropic region with
coexistence of rhombic (pseudo-cubic) and tetragonal symmetry phases, its
concentration being dependent on temperature. The maximal contribution of
pseudo-cubic symmetry polarization at $x=0.135$ was supposed to be related
to high values of electromechanical coupling coefficient, observed at $%
x\approx 0.1$ in the mixed system. The developed theory describes pretty
good observed phase diagram of PSN-PST and PMN-PT mixed relaxor systems.

\end{multicols}

\end{document}